\documentstyle[preprint,aps,version2]{revtex}
\begin{document}
\draft
\begin{title}
{\bf GAUGE INVARIANT EFFECTIVE POTENTIAL FOR\\
     ABELIAN MAXWELL--CHERN--SIMONS SYSTEMS}
\end{title}

\author{D. Bazeia and J. R. S. Nascimento}

\begin{instit}
Departamento de F\'\i sica, Universidade Federal da Para\'\i ba\\Caixa Postal 5008, 58051-970 Jo\~ao Pessoa, Para\'\i ba, Brazil
\end{instit}

\begin{abstract}
We investigate the effective potential for Abelian Maxwell--Chern--Simons systems. The calculations follow an alternate approach, recently proposed as a gauge invariant formulation of the effective potential, constructed in terms of a gauge invariant order parameter. We compare the results with another investigation, obtained within a standard route of calculating the effective potential.
\end{abstract}



This paper deals with the effective potential in Abelian gauge theories that contain the Chern--Simons term in the action of the gauge field, in tridimensional spacetime. 
As we know, quantum and thermal corrections to the classical potential in Abelian Chern--Simons and Maxwell--Chern--Simons systems were already obtained \cite{ilt92}. Another investigation was also performed \cite{baz93}, in this case addressing the issue concerning gauge invariance of the effective potential in Abelian Chern--Simons and Maxwell--Chern--Simons theories.

Despite these investigations, in the present paper we
return to issues concerning the effective potential in tridimensional Chern--Simons systems. The novelty here is the program very recently introduced in \cite{bbh96}, offered as an alternate procedure to obtain gauge invariant effective potentials. Our main motivation is to extend this approach to planar Chern--Simons systems. The key interest of the present investigation is to obtain the zero-temperature one-loop quantum corrections to the classical potential, and compare this result with another one, obtained via a standard procedure of calculating the effective potential.

In a former investigation \cite{baz93}, the issue concerning
gauge invariance of the effective potential was addressed within a standard route \cite{nie75,afr84,nba87,lba90}. The explicit calculations there presented have shown that the gauge invariant features of the effective potential
remain unchanged when one goes from $(3+1)$ spacetime dimensions to the tridimensional Chern--Simons territory. Hence we expect that the present investigation will not
change the qualitative features already presented in Ref.~{\cite{bbh96}}. And this is one of the results we shall explicitly offer in this work.

The tridimensional system that we consider is defined by the Lagrangian density
\begin{equation}
\label{eq:system}
{\cal L}=-\frac{1}{4}F_{\mu\nu}F^{\mu\nu}+\frac{1}{4}\kappa
\epsilon^{\mu\nu\lambda}A_{\mu}F_{\nu\lambda}+
|D_{\mu}\varphi|^2-V(|\varphi|).
\end{equation}
The two first terms in the above expression describes Abelian Maxwell--Chern--Simons dynamics, and we are using standard notation, with the covariant derivative given by
$D_{\mu}=\partial_{\mu}+ieA_{\mu}$. The potential is polynomial, and can be of up to sixth order in $|\varphi|$.
It can have the following forms, for instance,
\begin{eqnarray}
V(|\varphi|)&=&\frac{1}{2}e^2(|\varphi|^2-a^2)^2\\ 
V(|\varphi|)&=&\frac{e^4}{\kappa^2}|\varphi|^2
(|\varphi|^2-a^2)^2.
\end{eqnarray}
The first potential corresponds to the case where self-dual vortex solutions appear when one removes the Chern--Simons term from the classical action \cite{bog76,vsh76}. The second potential reproduces the self-dual Chern--Simons system when one removes the Maxwell term from the classical action \cite{jwe90}. 

The procedures that lead to the reduced Maxwell or Chern--Simons dymanics go as follows: In the above Lagrangian density one first changes $A_{\mu}\to A_{\mu}/e$, and then set
$\kappa \to 0$, to get to the Maxwell system, or set $e^2\to\infty$, mantaining $\kappa/e^2$ finite, to get to the pure Chern--Simons system. In spite of this, here we shall leave the potential unspecified, although we require that it presents spontaneous symmetry breaking. We can perhaps have a better way of doing a general investigation in Maxwell--Chern--Simons systems. Here we refer to the case where we work with the enlarged Maxwell--Chern--Simons system \cite{llm90,baz91}. This system is interesting since it is self-dual and allows getting the self-dual Maxwell or Chern--Simons system via very nice limiting procedures. However, the key issue we shall be discussing in this paper has nothing to do with self-duality, and the investigation does not change, at least qualitatively, even when one discards the Maxwell term from the classical action. Hence, we shall keep working with the Lagrangian density $(\ref{eq:system})$ to offer another explicit and illustrative example of the procedure introduced in Ref

We now follow the program of Ref.~{\cite{bbh96}}. In this case, the above Lagrangian density $(\ref{eq:system})$ presents the following canonical momenta conjugate to the vector and scalar fields 
\begin{eqnarray}
\Pi^0&=&0 \\
\Pi_i&=&\partial^0A^i-\partial^iA^0+\frac{1}{2}
\kappa\epsilon^{ij}A^j \\
\pi&=&\partial^0\varphi^{\dag}-ieA^0\varphi^{\dag} \\
\pi^{\dag}&=&\partial^0\varphi+ieA^0\varphi.
\end{eqnarray}
The Hamiltonian is then given by
\begin{eqnarray}
H&=&\int d^2x \Bigl\{ \frac{1}{2} \vec{\Pi}\cdot\vec{\Pi}+
\frac{1}{2}\kappa\vec{\Pi}\times\vec{A}+\frac{1}{2}
(\nabla\times\vec{A})^2+\frac{1}{8}\kappa^2\vec{A}\cdot\vec{A}+\nonumber\\
& &\qquad
\pi^{\dag}\pi+(\nabla\varphi-ie\vec{A}\varphi)\cdot(\nabla\varphi^{\dag}+ie\vec{A}\varphi^{\dag})+V(|\varphi|)+\nonumber\\
& &\qquad
A^0[\nabla\cdot\vec{\Pi}-\frac{1}{2}\kappa\nabla\times\vec{A}-ie(\pi\varphi-\pi^{\dag}\varphi^{\dag})]\Bigr\}.
\end{eqnarray}

We follow a standard route for dealing with constrained systems, as presented in \cite{bbh96}. In this case, the above system presents two first-class constraints, namely
\begin{equation}
\Pi^0=0,
\end{equation}
and the Gauss law constraint
\begin{equation}
\nabla\cdot\vec{\Pi}-\frac{1}{2}\kappa\nabla\times\vec{A}-
\rho=0,
\end{equation}
where the charge density is given by
\begin{equation}
\rho=ie(\pi\varphi-\pi^{\dag}\varphi^{\dag}).
\end{equation}

We write the gauge field variables in longitudinal and transversal components:
\begin{equation}
\vec{A}=\vec{A}_l+\vec{A}_t,
\end{equation}
with
\begin{eqnarray}
\nabla\times\vec{A}_l&=&0\\
\nabla\cdot\vec{A}_t&=&0
\end{eqnarray}
and
\begin{equation}
\vec{\Pi}=\vec{\Pi}_l+\vec{\Pi}_t,
\end{equation}
with
\begin{eqnarray}
\nabla\times\vec{\Pi}_l&=&0\\
\nabla\cdot\vec{\Pi}_t&=&0.
\end{eqnarray}
Gauss' law then gives
\begin{equation}
\vec{\Pi}_l=\nabla_x\int d^2y\,G(x-y)\left(\rho+\frac{1}{2}
\kappa\nabla\times\vec{A}\right)(y).
\end{equation}
In the gauge invariant subspace of wave-functionals, which appears in the Schr\"odinger representation \cite{jac90} of quantized fields, the above Hamiltonian can be written in the form
\begin{eqnarray}
H&=&\int d^2x \Bigl\{ \frac{1}{2} \vec{\Pi}_t\cdot\vec{\Pi}_t
+\pi^{\dag}\pi+\frac{1}{2}(\nabla\times\vec{A})^2+
\frac{1}{8}\kappa^2\vec{A}_t\cdot\vec{A}_t+
\nonumber\\
& &\qquad(\nabla\varphi-ie\vec{A}_t\varphi)\cdot
(\nabla\varphi^{\dag}+ie\vec{A}_t\varphi^{\dag})+V(|\varphi|)+
\nonumber\\
& &\qquad\frac{1}{2}\int d^2y [\rho(x)+2\kappa
(\nabla\times\vec{A})(x)]G(x-y)\rho(y)
\Bigr\}.
\end{eqnarray}
This Hamiltonian is gauge invariant, and for $\kappa\to 0$ it reproduces the result obtained in \cite{bbh96}, apart from the potential and an extra space coordinate in the case there investigated. In order to better understand the above result, we refer the reader to the reasoning presented in Sect.~{II}
of Ref.~{\cite{bbh96}}, and we also quote Ref.~{\cite{jac90}}, in which the Shr\"odinger representation for quantized fields is nicely presented, including the case of planar Chern--Simons theories.

To calculate the effective potential, let us now consider the scalar field as $\varphi=\chi+\phi$ and $\varphi^{\dag}=\chi^{\dag}+\phi^{\dag}$, where $\chi$ is the classical, constant piece, and $\phi$ represents the first quantum corrections. We collect the classical and the bilinear
quantum corrections to the Hamiltonian to get
\begin{eqnarray}
H&=&\Omega V(|\chi|)+\int d^2x \Biggl\{ \frac{1}{2} \vec{\Pi}_t\cdot\vec{\Pi}_t+
\frac{1}{2}(\nabla\times\vec{A})^2+
\nonumber\\
& &\left(e^2|\chi|^2+\frac{1}{2}\kappa^2\right)
\vec{A}_t\cdot\vec{A}_t+
\pi^{\dag}\pi+\nabla\phi^{\dag}\cdot\nabla\phi+
V_b(|\varphi|)+\nonumber\\
& &\frac{1}{2}\int d^2y [\bar{\rho}(x)+
2\kappa(\nabla\times\vec{A})(x)]G(x-y)\bar{\rho}(y)\Biggr\},
\end{eqnarray}
where $\Omega$ is the area of the planar space. Here we have set $\bar{\rho}=ie(\pi\chi-\pi^{\dag}\chi^{\dag})$, and $V_b(|\phi|)$ represents the bilinear contributions that appear from the potential. We introduce
\begin{eqnarray}
\phi&=&2^{-1/2}(\phi_1+i\phi_2)\\
\pi&=&2^{-1/2}(\pi_1-i\pi_2)
\end{eqnarray}
to write the bilinear terms in the Hamiltonian in the form
\begin{equation}
H_b=\frac{1}{2}\int d^2x[\pi_1\pi_1+\phi_1(-\nabla^2+m_H^2)\phi_1 ]+H'_b
\end{equation}
where $m^2_H=m^2_H(|\chi|)$ stand for the quadratic $\phi_1^2$ contributions that comes from the potential, and $H'_b$ is given by
\begin{eqnarray}
H'_b&=&\frac{1}{2}\int d^2x \Bigl\{ \Pi_t^2+A_t^i(-\nabla^2+2e^2|\chi|^2+\kappa^2)A_t^i+
\nonumber\\
& &\qquad \pi_2\pi_2+\phi_2(-\nabla^2+m_G^2)\phi_2 +\nonumber\\
& &\qquad\int d^2y\,[\bar{\rho}(x)+2\kappa(\nabla\times\vec{A})(x)]
G(x-y)\bar{\rho}(y)
\Bigr\},
\end{eqnarray}
where $m^2_G=m^2_G(|\chi|)$ represents the quadratic $\phi^2_2$ contributions that comes from the potential. 

Let us now go to momenta space. In this case we get
\begin{equation}
H_b=\frac{1}{2}\int \frac{d^2k}{(2\pi)^2} [\pi_1(k)\pi_1(-k)+w_H^2\phi_1(k)\phi_1(-k)]+H'_b,
\end{equation}
where
\begin{equation}
w_H^2=k^2+m_H^2,
\end{equation}
and $H'_b$ is now given by
\begin{eqnarray}
H'_b&=&\frac{1}{2}\int\frac{d^2k}{(2\pi)^2}\Biggl\{ \Pi_t^i(k)\Pi_t^i(-k)+w_t^2(k)A_t^i(k)A_t^i(-k)+\nonumber\\
& & \pi_2(k)\pi_2(-k)+w_G^2(k)\phi_2(k)\phi_2(-k) +\nonumber\\
& &{2i2^{1/2}e\kappa\chi}\epsilon^{ij}
\frac{k^i}{k^2}A^j(k)\pi_2(-k)+
\frac{w_A^2}{k^2}\pi_2(k)\pi_2(-k)\Biggr\},
\end{eqnarray}
where we have set
\begin{eqnarray}
w_T^2&=&k^2+2e^2|\chi|^2+\kappa^2\\
w_G^2&=&k^2+m_G^2\\
w_A^2&=&k^2+2e^2|\chi|^2
\end{eqnarray}

To obtain the first quantum corrections, the one-loop contributions, we can proceed as follows: We write the bilinear contributions to the Hamiltonian in terms of a new field $\Phi$, which can be written as
\begin{equation}
\Phi^{\dag}=(\Pi^i_t, A^i_t,\pi_2,\phi_2).
\end{equation}
The bilinear contributions to the Hamiltonian can be
rewritten as
\begin{equation}
H'_b=\frac{1}{2}\int \frac{d^2k}{(2\pi)^2}\,\Phi_i^{\dag}M_{ij}\Phi_j,
\end{equation}
where $M$ is a matrix that can be written as
\begin{equation}\left(
\begin{array}{cccc}
1 & 0&0 &0 \\
0&w_T^2&i2^{1/2}e\kappa|\chi|\epsilon^{ij}\frac{k^j}{k^2}&0\\
0&-i2^{1/2}e\kappa|\chi|\epsilon^{ij}\frac{k^j}{k^2}&
\frac{w_A^2}{k^2}&0\\
0 &0&0 &w_G^2 \end{array}\right)
\end{equation}
The effective potential is then given by, at the one-loop level,
\begin{eqnarray}
\label{eq:result1}
V^1(|\chi|)&=&V(|\chi|)+\frac{1}{2}\int\frac{d^2k}{(2\pi)^2}
\Biggl[\sqrt{k^2+m_H^2}+\nonumber\\
& &\qquad\sqrt{k^2+m_G^2}+\sqrt{k^2+m_{+}^2}+
\sqrt{(k^2+m_{-}^2}\Biggr].
\end{eqnarray}
Here we have dropped irrelevant divergent contributions, and we have set
\begin{equation}
m^2_{\pm}=2e^2|\chi|^2+\frac{1}{2}\kappa^2\pm
\frac{1}{2}\kappa^2[1+8e^2|\chi|^2/\kappa^2]^{1/2}.
\end{equation}
We notice that the limit $\kappa\to 0$, which removes the Chern--Simons term, reproduces the result obtained in Ref.~{\cite{bbh96}} -- recall that here we are working with planar systems.

Let us now compare the above result with a former result, obtained in \cite{baz93} via a standard approach. We
recall that the Maxwell--Chern--Simons system $(\ref{eq:system})$ presents the following one-loop effective potential \cite{baz93}
\begin{equation}
V_{R}^1(\bar{\phi})=V(\bar{\phi})+\frac{1}{2}
\int\frac{d^3k}{(2\pi)^3}
[\ln(w_{H}^2\,w_{+}^2\,w_{-}^2\,
w_{\xi\upsilon}^2)-2\ln(w_{\upsilon}^2)].
\end{equation}
Here we are considering the classical $\bar{\phi}$ field real, as in Ref.~{\cite{baz93}}. Furthermore, we are using
\begin{eqnarray}
w_H^2&=&k^2+M_H^2\\
w_{\pm}^2&=&k^2+M^2_{\pm}\\
w_{\xi\upsilon}^2&=&(k^2+e^2\upsilon\bar{\phi})^2+
M_G^2(k^2+\xi e^2\bar{\phi}^2)\nonumber\\
&=&(k^2+M_{g+}^2)(k^2+M_{g-}^2)\\
w_v^2&=&(k^2+e^2\upsilon\bar{\phi}),
\end{eqnarray}
where $M_H^2$ and $M_G^2$ represents the quadratic $\phi_1^2$ and $\phi_2^2$ contributions that comes from the potential, respectively, and
\begin{eqnarray}
M_{\pm}^2&=&e^2\phi^2+\frac{1}{2}\kappa^2\pm
\frac{1}{2}\kappa^2[1+4e^2\phi^2/\kappa^2]^{1/2}\\
M_{g\pm}^2&=&{\frac{1}{2}M_G^2+e^2\upsilon\bar{\phi}}\pm
\frac{1}{2}{M_G^2}{\Bigl[1+
8e^2\bar{\phi}(\upsilon-\xi\bar{\phi})/M_G^2\Bigr]}^{1/2}.
\end{eqnarray}

In the above result the terms $M^2_{g\pm}$ are gauge-dependent contributions that appear from the procedure \cite{baz93} which considers the general $R$ gauge. In this case $\xi$ and $\upsilon$ are the two gauge parameters required by the gauge choice.

In order to compare this result with the former result $(\ref{eq:result1})$, obtained within a Hamiltonian approach, let us now integrate the above expression over the time component in momentum space. Here we get, after dropping irrelevant divergent contributions,
\begin{eqnarray}
V_{R}^1(\phi)&=&V(\phi)+\frac{1}{2}\int\frac{d^2k}{(2\pi)^2}
\Biggl[\sqrt{k^2+M_H^2}+\sqrt{k^2+M_{+}^2}+\nonumber\\
& &\sqrt{k^2+M_{-}^2}+
\sqrt{k^2+M_{g+}^2}+\sqrt{k^2+M_{g-}^2}-
2\sqrt{k^2+e^2\upsilon\bar{\phi}}\Biggr].
\end{eqnarray}
In $R_{\xi}$ gauge we must set $\upsilon\to\xi\bar{\phi}$. In this case we get
\begin{eqnarray}
V_{R_{\xi}}^1(\bar{\phi})&=&V(\bar{\phi})+\frac{1}{2}
\int\frac{d^2k}
{(2\pi)^2}\Biggl[\sqrt{k^2+M_H^2}+\sqrt{k^2+M_{+}^2}+
\nonumber\\
& &\sqrt{k^2+M_{-}^2}+\sqrt{k^2+M_G^2+\xi e^2\bar{\phi}^2}- \sqrt{k^2+\xi e^2\bar{\phi}^2}\Biggr].
\end{eqnarray}
In Landau gauge we now set $\xi\to 0$ to obtain, after discarding an irrelevant (divergent) contribution,
\begin{eqnarray}
\label{eq:result2}
V_{L}^1(\bar{\phi})&=&V(\bar{\phi})+\frac{1}{2}\int
\frac{d^2k}{(2\pi)^2}\Biggl[\sqrt{k^2+M_H^2}+\nonumber\\
& &\qquad\sqrt{k^2+M_G^2}+\sqrt{k^2+M_{+}^2}+
\sqrt{k^2+M_{-}^2}\Biggr].
\end{eqnarray}
We notice that the above result reproduces the result given 
by Eq.~{(\ref{eq:result1})}, obtained within the program introduced in \cite{bbh96}. In fact, there is a slight difference between the results $(\ref{eq:result1})$ and 
$(\ref{eq:result2})$, which disappears when one
sets $\bar{\phi}\to 2^{1/2}|\chi|$. Here we recall that in the approach of Ref.~{\cite{bbh96}}, the complex field $\chi$ appears as a gauge invariant quantity, transforming with a constant phase under the $U(1)$ symmetry, whereas in the standard Faddeev--Popov path-integral procedure one usually shifts the complex field by a real piece, here represented by 
$\bar{\phi}$.

To end this paper, let us now recall that we have extended 
the approach presented in \cite{bbh96} to the case of planar Chern--Simons systems. As we have shown, the presence of the Chern--Simons term changes the constraints of the system, but the constrained Maxwell-Chern--Simons system is qualitatively similar to the standard Maxwell system, and hence the program offered in Ref.~{\cite{bbh96}} can be implemented without further difficulty.

\acknowledgments
We would like to thank F. A. Brito and R. F. Ribeiro for several interesting discussions.

\end{document}